\documentstyle[aps,epsf,manuscript]{revtex}

\begin{document}

\newcommand{\half}{\frac12}
\newcommand{\vare}{\varepsilon}
\newcommand{\eps}{\epsilon}
\newcommand{\pr}{^{\prime}}
\newcommand{\ppr}{^{\prime\prime}}
\newcommand{\pp}{{p^{\prime}}}
\newcommand{\hp}{\hat{\bfp}}
\newcommand{\hpp}{\hat{\bfpp}}
\newcommand{\hq}{\hat{\bfq}}
\newcommand{\rqq}{{\rm q}}
\newcommand{\rx}{{\rm x}}
\newcommand{\rp}{{\rm p}}
\newcommand{\rpp}{{{\rm p}^{\prime}}}
\newcommand{\rk}{{\rm k}}
\newcommand{\bfp}{{\bf p}}
\newcommand{\bfpp}{{\bf p}^{\prime}}
\newcommand{\bfq}{{\bf q}}
\newcommand{\bfx}{{\bf x}}
\newcommand{\bfk}{{\bf k}}
\newcommand{\bfz}{{\bf z}}
\newcommand{\bphi}{{\mbox{\boldmath$\phi$}}}
\newcommand{\balpha}{{\mbox{\boldmath$\alpha$}}}
\newcommand{\bsigma}{{\mbox{\boldmath$\sigma$}}}
\newcommand{\bomega}{{\mbox{\boldmath$\omega$}}}
\newcommand{\bvare}{{\mbox{\boldmath$\varepsilon$}}}
\newcommand{\intzo}{\int_0^1}
\newcommand{\intinf}{\int^{\infty}_{-\infty}}
\newcommand{\ka}{\kappa_a}
\newcommand{\kb}{\kappa_b}
\newcommand{\lbr}{\langle}
\newcommand{\rbr}{\rangle}
\newcommand{\ThreeJ}[6]{
        \left(
        \begin{array}{ccc}
        #1  & #2  & #3 \\
        #4  & #5  & #6 \\
        \end{array}
        \right)
        }
\newcommand{\SixJ}[6]{
        \left\{
        \begin{array}{ccc}
        #1  & #2  & #3 \\
        #4  & #5  & #6 \\
        \end{array}
        \right\}
        }
\newcommand{\NineJ}[9]{
        \left\{
        \begin{array}{ccc}
        #1  & #2  & #3 \\
        #4  & #5  & #6 \\
        #7  & #8  & #9 \\
        \end{array}
        \right\}
        }
\newcommand{\Dmatrix}[4]{
        \left(
        \begin{array}{cc}
        #1  & #2   \\
        #3  & #4   \\
        \end{array}
        \right)
        }
\newcommand{\cross}[1]{#1\!\!\!/}
\newcommand{\beq}{\begin{equation}}
\newcommand{\eeq}{\end{equation}}
\newcommand{\beqn}{\begin{eqnarray}}
\newcommand{\eeqn}{\end{eqnarray}}

%
%
\title{
Two-loop self-energy correction in H-like ions
}
\author{
V.~A.~Yerokhin and  V.~M.~Shabaev
}

\address{
Department of Physics, St. Petersburg State University, Oulianovskaya 1, Petrodvorets,
St. Petersburg 198504, Russia }
\date{\today}
\maketitle

\begin{abstract}
A part of the two-loop self-energy correction, the so-called P term, is evaluated
numerically for the $1s$ state to all orders in $Z\alpha$. Our calculation, combined
with the previous investigation [S. Mallampalli and J. Sapirstein, Phys. Rev. A {\bf
57}, 1548 (1998)], yields the total answer for the two-loop self-energy correction in
H-like uranium and bismuth. As a result, the major uncertainty is eliminated from the
theoretical prediction for the Lamb shift in these systems. The total value of the
ground-state Lamb shift in H-like uranium is found to be 463.93(50) eV.

\noindent PACS number(s): 31.30.Jv, 31.10.+z
\end{abstract}

%
%
\section*{Introduction}

The calculation of all two-loop QED diagrams for the Lamb shift of H-like ions is one
of the most challenging problems in bound-state QED. The experimental accuracy
of 1 eV aimed at in measurements of the ground-state Lamb shift in H-like uranium
\cite{Stoehlker00} requires a calculation of the complete set of QED corrections of the
order $\alpha^2$ without any expansion in the parameter $Z\alpha$ ($Z$ is the nuclear
charge number, $\alpha$ is the fine structure constant). In high-$Z$ Li-like ions,
these diagrams are the source of the major theoretical uncertainty for the
$2p_{1/2}$-$2s$ transition energy \cite{Yerokhin01} and, therefore, the limiting
factor in comparison of theory and experiment. Also in the low-$Z$ region, the
two-loop Lamb shift is important from the experimental point of view
\cite{Pachucki01}. What is more, its $Z\alpha$ expansion exhibits a rather peculiar
behavior, with a very slow convergence even in case of hydrogen \cite{Pachucki01}. In
order to eliminate the uncertainty due to higher-order contributions, it is important
to perform a non-perturbative (in $Z\alpha$) calculation of two-loop corrections
even in the low-$Z$ region.

The most problematic part of the one-electron $\alpha^2$ contribution is the two-loop
self-energy correction, represented diagramatically in Fig.~\ref{sese}. The diagram
in Fig.~\ref{sese}(a) is usually divided into two parts, which are referred to as
the irreducible and the reducible contribution. (The reducible contribution is
defined as a part of this diagram in which intermediate states in the spectral
decomposition of the middle electron propagator coincide with the initial state.) The
irreducible contribution (also referred to as the loop-after-loop correction) can be
shown to be gauge invariant when covariant gauges are used. Its evaluation is not very
cumbersome and was accomplished in several independent investigations
\cite{Mitrushenkov95,Mallampalli98a,Yerokhin00}.

The reducible part of the diagram in Fig.~\ref{sese}(a) should be evaluated together
with the remaining two diagrams in Fig.~\ref{sese}. This calculation is by far more
difficult. The first attempt to attack this problem was performed by Mallampalli and
Sapirstein \cite{Mallampalli98}. In that work, the contribution of interest was
rearranged into three parts (referred to by the authors as the M, P, and F terms), only
two of which were actually evaluated. The remaining part (the P term) was left out
since a new  numerical technique had to be developed for its computation. In our
present investigation, we perform the numerical evaluation of the missing part of the
two-loop self-energy, the P term. Results of our evaluation, added to those from
Ref.~\cite{Mallampalli98}, yield the final answer for the total two-loop self-energy
correction for H-like uranium and bismuth. This result disagrees with the resent
calculation of the total contribution reported by Goidenko {\it et al.}
\cite{Goidenko01}, which is based on the partial-wave renormalization approach.

The plan of the paper is the following. The basic formulas needed for the evaluation
of the P term are given in the first section, alongside with the discussion of the
treatment of ultraviolet and infrared divergences. In the next section we formulate
the scheme of the numerical evaluation and give some technical details. Numerical
results are discussed in the last section. In that section we also collect all
second-order contributions to the ground-state Lamb shift of H-like uranium and to the
$2p_{1/2}$-$2s$ transition energy in Li-like uranium. In the latter case, the two-loop
self-energy contribution is estimated by scaling the $1s$ result.

In this paper, we use the relativistic units ($\hbar=c=m=1$).
The roman style ($\rp$) is used for four vectors, bold face ($\bfp$) for three
vectors and italic style ($p$) for scalars.
Four vectors have the form $\rp \equiv (\rp_0,\bfp)$.
The scalar product of two four vectors is
$(\rp \cdot \rk) = \rp_0 \rk_0 - (\bfp \cdot \bfk)$.
We use the notations $\cross{\rp} = {\rm p}_{\mu} \gamma^{\mu}$,
$\hat{\bfp} = \bfp/|\bfp|.$

%
\section{Basic formalism}

In this paper we are concerned with the evaluation of the correction
\beq                                            \label{bf1}
\Delta E = \Delta E^{N1}+ \Delta E^{N2}+ 2 \Delta E^{O1} \ ,
\eeq
where the contributions $\Delta E^{N1}$, $\Delta E^{N2}$, and $\Delta E^{O1}$ are
represented by Feynman diagrams shown in Figs.~\ref{dEN1}, \ref{dEN2}, and
\ref{dEO1}, respectively. Our consideration of these three sets of Feynman diagrams
should be considered as an investigation, complementary to Ref. \cite{Mallampalli98},
to which we refer hereafter as I. In our calculation we use the Feynman gauge and the
point nuclear model, the same as in I. Our results, combined with those from I, should
yield the complete answer for the two-loop self-energy correction.

All the contribution $\Delta E^{N1}$, $\Delta E^{N2}$, and $\Delta E^{O1}$ are
ultraviolet (UV) divergent. Following I, we refer to their UV-finite part  as the ``P
term''. We note that subtractions in these contributions are chosen in such a way
that each of them is free from overlapping divergences. The main problem in the
evaluation of these diagrams is that they contain bound electron propagators as well
as UV divergences. While UV divergences are normally separated in momentum space, the
Dirac-Coulomb Green function is generally treated in the coordinate representation.
The most direct way for the calculation of the P term consists in developing a
numerical scheme for the evaluation of the Dirac Coulomb Green function in momentum
space, which is one of the aims of the present work.

The nested contributions, $\Delta E^{N1}$ and $\Delta E^{N2}$, possess in addition
some infrared (IR) divergences, which are associated with the so-called
reference-state singularities. These divergences are cancelled out when considered
together with the corresponding parts of the reducible contribution of the diagram in
Fig.~\ref{sese}(a). Following I, we handle the IR divergences by introducing a
regulator. This makes clear that great care should be taken in order to separate all
divergences exactly in the same way as in I, in order not to miss a finite
contribution.

\subsection{One-loop self-energy}

We start with some basic formulas for the first-order
self-energy correction. The formal expression for the
unrenormalized first-order self-energy matrix element in the
Feynman gauge is given by

\beqn \label{eq1}
\Delta E^{(1)}_{\rm unren} &=&
\frac{i\alpha}{2\pi} \intinf d\omega \int d\bfx_1
        d \bfx_2 \, \psi^{\dag}_a(\bfx_1) \alpha_{\mu}
        G(\vare_a-\omega,\bfx_1,\bfx_2) \alpha^{\mu}
        \psi_a(\bfx_2) \frac{\exp(i|\omega|x_{12})}{x_{12}}
\ ,
\eeqn
where $ {\alpha}^{\mu} = (1, \balpha) $ are the Dirac matrices,
and $G$ is the Dirac-Coulomb Green function,
\beq \label{eq2}
G(\vare,\bfx_1,\bfx_2) = \sum_n
        \frac{ \psi_n(\bfx_1) \psi^{\dag}_n(\bfx_2)}
                {\vare-\vare_n(1-i0)} \ .
\eeq
Eq.~(\ref{eq1}) is written completely in coordinate space. We will need also the
corresponding expression in the mixed momentum-coordinate representation. This can be
obtained by the Fourier transformation of Eq.~(\ref{eq1}) over one of the radial
variables,
\beqn                                           \label{eq3}
\Delta E^{(1)}_{\rm unren} &=&
-\frac{i\alpha}{2\pi} \intinf d\omega \int \frac{d\bfp_1}{(2\pi)^3}
        \frac{d\bfp_2}{(2\pi)^3} \sum_n
         \frac{A_{\mu}^{an}(\omega,\bfq) \psi_n^{\dag}(\bfp_1)
                \alpha^{\mu}\psi_a(\bfp_2)}
                        {\vare_a-\omega-\vare_n(1-i0)} \ ,
\eeqn
\beqn                                            \label{eq4}
A_{\mu}^{an}(\omega,\bfq) &=&
 \frac{4\pi}{\omega^2-\bfq^2+i0} \int d\bfx \, \psi_a^{\dag}(\bfx)
        \alpha_{\mu} \psi_n(\bfx) e^{-i\bfq \bfx} \ ,
\eeqn
where $\bfq = \bfp_1-\bfp_2$, and
\begin{equation}                                \label{eq5}
\psi(\bfp) = \int d \bfx  \, e^{-i\bfp\bfx}\psi(\bfx) \ .
\end{equation}
We note that while the integration over $\omega$ in Eq.~(\ref{eq3}) can be carried out
by Cauchy's theorem, we prefer to keep it, having in mind future generalizations to
the two-loop case.

The renormalization of the one-loop self-energy is well known. In our work, we employ
the method based on the expansion of the bound electron propagator in Eq.~(\ref{eq1})
in terms of the interaction with the nuclear Coulomb field \cite{Snyderman91}. For the
detailed description of our renormalization procedure we refer the reader to
\cite{Yerokhin99}. The renormalized self-energy correction is represented by the sum
of three finite terms,
\beq                                            \label{eq6}
\Delta E_{\rm ren}^{(1)} = \Delta E^{(1)}_{\rm zero}+
        \Delta E^{(1)}_{\rm one}+ \Delta E^{(1)}_{\rm many} \ ,
\eeq
where
\begin{equation}                                \label{eq7}
\Delta E_{\rm zero}^{(1)} = \int \frac{d\bfp}{(2\pi)^3} \,
        \psi^{\dag}_a(\bfp) \gamma_0 \Sigma^{(0)}_R(\vare_a,\bfp) \psi_a(\bfp) \ ,
\end{equation}
\begin{equation}                                 \label{eq8}
\Delta E_{\rm one}^{(1)} =
        \int \frac{d\bfp_1}{(2\pi)^3}
         \frac{d\bfp_2}{(2\pi)^3} \,
        \psi^{\dag}_a(\bfp_1) \gamma_0 \Gamma^0_R(\vare_a,\bfp_1;\vare_a,\bfp_2)
        V_C(\bfq) \psi_a(\bfp_2) \ ,
\end{equation}
where  $V_C(\bfq) = -4\pi\alpha Z/|\bfq|^2$, and $\Sigma^{(0)}_R(\rp)$ and
$\Gamma^{\mu}_R(\rp_1,\rp_2)$ are the renormalized free self-energy and vertex
operators introduced in Appendix~\ref{ap1}. The expression for $\Delta E^{(1)}_{\rm
many}$ is given by Eq.~(\ref{eq1}), where the Green function
$G(\vare_a-\omega,\bfx_1,\bfx_2)$ is replaced with
$G^{(2+)}(\vare_a-\omega,\bfx_1,\bfx_2)$,
\beq                                            \label{eq9}
G^{(2+)}(\vare,\bfx_1,\bfx_2) =
    G(\vare,\bfx_1,\bfx_2)
 -  G^{(0)}(\vare,\bfx_1,\bfx_2)-
    G^{(1)}(\vare,\bfx_1,\bfx_2) \ ,
\eeq
$G^{(0)}(\vare,\bfx_1,\bfx_2)$ is the free Dirac Green function, and
\beq                                            \label{eq10}
    G^{(1)}(\vare,\bfx_1,\bfx_2) =
        -\int d\bfz G^{(0)}(\vare,\bfx_1,\bfz) \frac{\alpha Z}{|\bfz|}
                G^{(0)}(\vare,\bfz,\bfx_2) \ .
\eeq

\subsection{Basic formulas for two-loop diagrams}

\begin{figure}
\centerline{ \mbox{ \epsfysize=5cm \epsffile{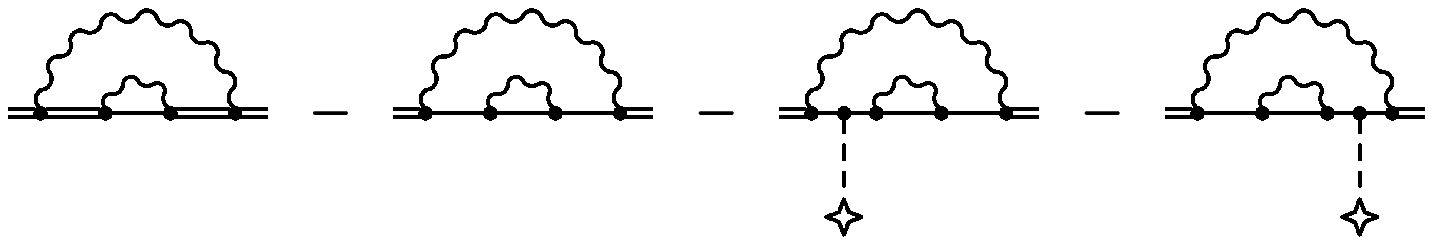} }} \caption{\label{dEN1}
Diagramatic representation of the correction $\Delta E^{N1}$. For brevity, we do not
explicitly display the diagrams involving mass conterterms; the inner self-energy loop
should be understood with the corresponding mass counterterm subtracted. }
\end{figure}

The formal expression for the first diagram in Fig.~\ref{dEN1} can be obtained from
Eq.~(\ref{eq1}) by the substitution
$
G(\vare_a-\omega,\bfx_1,\bfx_2) \to
        {G}_{N1}(\vare_a-\omega,\bfx_1,\bfx_2) \ ,
$
where
\beqn                                           \label{bf2}
{ G}_{N1}(\vare,\bfx_1,\bfx_2) &=&
        \int \frac{d\bfp}{(2\pi)^3} \, G(\vare,\bfx_1,\bfp)
           \gamma_0 \biggl[ \Sigma^{(0)}(\vare,\bfp)
              -\delta m \biggr]
                        G(\vare,\bfp,\bfx_2) \ ,
\eeqn
\beqn                                           \label{bf2a}
G(\vare,\bfx_1,\bfp) &=&
        \int d\bfx_2 \, e^{i\bfp \bfx_2} G(\vare,\bfx_1,\bfx_2) \ ,
            \\                                  \label{bf2b}
G(\vare,\bfp,\bfx_2) &=&
        \int d\bfx_1 \, e^{-i\bfp \bfx_1} G(\vare,\bfx_1,\bfx_2) \ ,
\eeqn
and $\Sigma^{(0)}(\rp)$ is the free one-loop self-energy operator defined in
Appendix~\ref{ap1}.

\begin{figure}
\centerline{ \mbox{
\epsfysize=5cm \epsffile{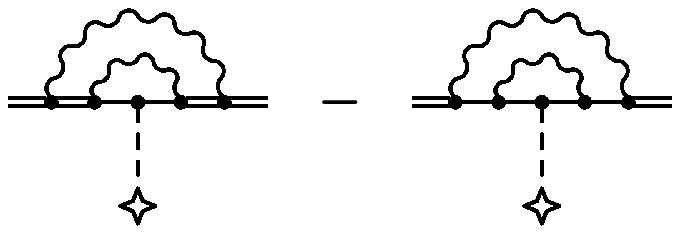}
}}
\caption{\label{dEN2}
Diagrammatic representation of the correction $\Delta E^{N2}$.
}
\end{figure}

The expression for the first diagram in Fig.~\ref{dEN2} is obtained from
Eq.~(\ref{eq1}) by the replacement
$
G(\vare_a-\omega,\bfx_1,\bfx_2) \to
        { G}_{N2}(\vare_a-\omega,\bfx_1,\bfx_2) \ ,
$
where
\beqn                                           \label{bf3}
{ G}_{N2}(\vare,\bfx_1,\bfx_2) &=&
        \int \frac{d\bfp_1}{(2\pi)^3} \frac{d\bfp_2}{(2\pi)^3} \,
                 G(\vare,\bfx_1,\bfp_1)
                        V_C(\bfq) \gamma_0
                \Gamma^0(\vare,\bfp_1;\vare,\bfp_2)
                        G(\vare,\bfp_2,\bfx_2) \ ,
\eeqn
where $\Gamma^{\mu}(\rp_1,\rp_2)$ is the free one-loop vertex operator defined in
Appendix~\ref{ap1}, and $V_C$ is the Coulomb potential in momentum space.

\begin{figure}
\centerline{ \mbox{
\epsfysize=5cm \epsffile{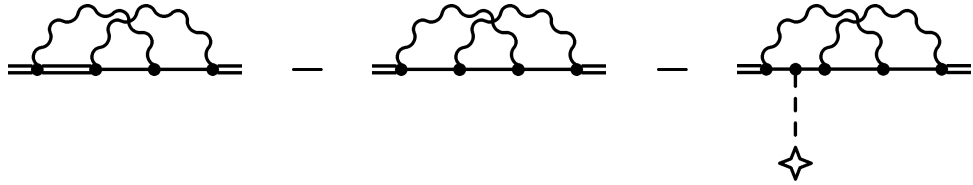}
}}
\caption{\label{dEO1}
Diagrammatic representation of the correction $\Delta E^{O1}$.
}
\end{figure}

We obtain the expression for the first diagram in Fig.~\ref{dEO1} by replacing one of
the $\gamma$ matrices in Eq.~(\ref{eq3}) with the vertex operator $\Gamma$,
\beqn                                           \label{bf4}
\Delta E^{O1}_{\rm unren} &=&
-\frac{i\alpha}{2\pi} \intinf d\omega \int \frac{d\bfp_1}{(2\pi)^3}
        \frac{d\bfp_2}{(2\pi)^3} \sum_n
         \frac{A_{\mu}^{an}(\omega,\bfq)}
                        {\vare_a-\omega-\vare_n(1-i0)}
                \nonumber \\ && \times
 \psi_n^{\dag}(\bfp_1)
        \gamma_0\Gamma^{\mu}(\vare_a-\omega,\bfp_1;\vare_a,\bfp_2)
                        \psi_a(\bfp_2)
\ ,
\eeqn
with $A_{\mu}^{an}(\omega,\bfq)$ given by Eq.~(\ref{eq4})

The expressions for the remaining diagrams in Figs.
\ref{dEN1}--\ref{dEO1} are obtained in a similar way, by replacing
the Dirac-Coulomb Green function $G$ in Eqs. (\ref{bf2}), (\ref{bf3}),
and (\ref{bf4}) by $G^{(0)}$ or $G^{(1)}$.

\subsection{Separation of ultraviolet divergences}

In this section we isolate the UV-finite part of $\Delta E$. Following I, we refer to
this contribution as the P term. Considering the renormalization of the diagrams in
Fig.~\ref{dEN1}, we should keep in mind that the inner self-energy loops are always
accompanied by the corresponding mass counterterms.

The renormalization of the one-loop self-energy and vertex operators is defined in
Appendix~\ref{ap1}. To handle the UV divergences, we regularize them by working in $D
= 4-\epsilon$ dimensions. The resulting expressions are
\beqn                                   \label{bf5}
\Sigma^{(0)}(\rp)-\delta m &=& \tilde{B}^{(1)} (\cross{\rp}-m)
                        + \Sigma^{(0)}_R(\rp) \ , \\
\Gamma^{\mu}(\rp_1,\rp_2) &=& \tilde{L}^{(1)} \gamma^{\mu}
                        + \Gamma^{\mu}_R(\rp_1,\rp_2) \ ,
\eeqn
where $\tilde{B}^{(1)}$ and $\tilde{L}^{(1)}$ are UV-divergent
constants, and $\Sigma^{(0)}_R(\rp)$ and $\Gamma^{\mu}_R(\rp_1,\rp_2)$
are finite. According to the Ward identity, $\tilde{B}^{(1)} =
-\tilde{L}^{(1)}$.

For the renormalization of the two-loop self-energy diagrams, we refer the
reader to a (very pedagogical) description given by Fox and Yennie
\cite{FoxYennie73}. Applying their arguments to the diagrams in Figs.
\ref{dEN1}--\ref{dEO1}, we have
\beqn                                   \label{bf6}
\Delta E^{N1}+ \Delta E^{N2} &=&
        \tilde{B}^{(1)} \Delta E^{(1)}_{{\rm many},D}
        + \Delta E^{N1}_P+ \Delta E^{N2}_P \ , \\
2 \Delta E^{O1} &=& 2 \tilde{L}^{(1)} \Delta E^{(1)}_{{\rm many},D}
        + 2 \Delta E^{O1}_P \ ,
\eeqn
where the subscript $P$ means that the corresponding contribution is UV convergent,
and the subscript $D$ of $\Delta E^{(1)}_{{\rm many},D}$ indicates that this
correction should be evaluated in $D$ dimensions, keeping terms of order $\epsilon$.
(These terms yield a finite contribution when multiplied by divergent
renormalization constants.)

The resulting expression reads
\beq                                    \label{bf7}
\Delta E = \tilde{L}^{(1)} \Delta E^{(1)}_{{\rm many},D}
+ \Delta E^{N1}_P+ \Delta E^{N2}_P+ 2 \Delta E^{O1}_P \ .
\eeq
Here, the correction $\Delta E^{N1}_P$ can be obtained from the corresponding
expression for $\Delta E^{N1}$ by the replacement $\Sigma^{(0)}(\rp) \to
\Sigma^{(0)}_R(\rp)$, and the corrections $\Delta E^{N2}_P$ and $\Delta E^{O1}_P$ --
by the corresponding substitution $\Gamma^{\mu}(\rp_1,\rp_2) \to
\Gamma^{\mu}_R(\rp_1,\rp_2)$. Since all the P-terms are UV-convergent, in their
evaluation we can disregard terms of order $\epsilon$ in definition of
$\Sigma^{(0)}_R(\rp)$ and $\Gamma^{\mu}_R(\rp_1,\rp_2)$. We note also that the
UV-divergent part of $\Delta E$ separated in Eq.~(\ref{bf7}), exactly corresponds to
that in I.

\subsection{Separation of infrared divergences}  \label{infr}

Infrared divergences occur in the corrections $\Delta E^{N1}_P$ and $\Delta E^{N2}_P$
due to so-called reference-state singularities. They arise when the energy of
intermediate states in the spectral decomposition of both electron propagators in Eqs.
(\ref{bf2}), (\ref{bf3}) coincide with the energy of the reference state $vare_a$. As
shown, {e.g.}, in I, the divergent terms disappear when considered together with the
related contributions from the reducible part of the diagram in Fig.~\ref{sese}(a).
However, since we are going to evaluate the contributions $\Delta E^{N1}_P$ and
$\Delta E^{N2}_P$ separately, a proper regularization of the IR divergences is needed.
In order to preserve the compatibility of our results with those of I, we have to
employ exactly the same procedure for the regularization of the IR divergences.

Following I, we introduce in the IR-divergent parts of $\Delta E^{N1}_P$, $\Delta
E^{N2}_P$ a regulator $\delta$ which modifies the location of the reference-state pole
of the Green function, $\vare_a \to \vare_a(1+\delta)$. After that, we have for the
IR-divergent part of $\Delta E^{N1}_P$
\beqn                                           \label{bf8}
\Delta E^{N1}_{P, IR}(\delta) &=& \frac{i\alpha}{2\pi}
        \sum_{\mu_{\overline{a}}} \intinf d\omega \,
            \frac{\lbr \overline{a}|\Sigma^{(0)}_R(\vare_a-\omega)|\overline{a}\rbr}
                {(-\vare_a\delta-\omega+i0)^2}
   \lbr a\overline{a}|
        \frac{1-\balpha_1 \balpha_2}{x_{12}} e^{i|\omega|x_{12}}|
          \overline{a}a\rbr \ ,
\eeqn
where $\overline{a}$ denotes the electron with the energy $\vare_a$ and the
angular-momentum projection $\mu_{\overline{a}}$. In Appendix~\ref{Apse} we
demonstrate that $\Sigma^{(0)}_R(\vare_a-\omega)$ as a function of $\omega$ can be
analytically continued to the first quadrant starting from the right half of the real
$\omega$ axis, and to the third quadrant from the left half of the real
$\omega$ axis. Therefore, we can perform the Wick rotation of the $\omega$ integration
contour in Eq.~(\ref{bf8}),
\beqn                                           \label{bf9}
\Delta E^{N1}_{P, IR}(\delta) &=& -\frac{\alpha}{\pi}
            \sum_{\mu_{\overline{a}}} \Re
        \int_0^{\infty} d\omega \,
            \frac{\lbr \overline{a}|\Sigma^{(0)}_R(\vare_a-i\omega)|\overline{a}\rbr}
                    {(\vare_a\delta+i\omega)^2}
   \lbr a\overline{a}|
        \frac{1-\balpha_1 \balpha_2}{x_{12}} e^{-\omega x_{12}}|
          \overline{a}a\rbr \ .
\eeqn

Let us investigate the behavior of $\Delta E^{N1}_{P, IR}$ for small values of
$\delta$. Writing it in a compact form, we have
\beqn                                           \label{bf10}
\Delta E^{N1}_{P, IR}(\delta) &=& \Re \int_0^{\infty} d\omega \,
        \frac{f(\omega)}{(\vare_a\delta+i\omega)^2}
                \nonumber \\
 &=& \Re \int_0^{\infty} d\omega \,
        \frac{f(\omega)-f(0)}{(\vare_a\delta+i\omega)^2}
                \nonumber \\
 &=& \Re \bigl[ f\pr(0) \bigr] \ln \delta+ \mbox{\rm terms, regular
in $\delta$} \ .
\eeqn
Taking into account that
\beq                                            \label{bf11}
\Re \left. \frac{d}{d\omega}\right|_{\omega = 0}
        \left[  \sum_{\mu_{\overline{a}}}
          \lbr \overline{a}|\Sigma^{(0)}_R(\vare_a-i\omega)|\overline{a}\rbr
       \lbr a\overline{a}|
        \frac{1-\balpha_1 \balpha_2}{x_{12}} e^{-\omega x_{12}}|
          \overline{a}a\rbr
        \right]  = -\lbr a|\Sigma^{(0)}_R(\vare_a)|a\rbr \ ,
\eeq
we have
\beq                                            \label{bf12}
\Delta E^{N1}_{P, IR}(\delta) = \frac{\alpha}{\pi}
        \Delta E^{(1)}_{\rm zero} \ln \delta + \Delta E^{N1}_{P,\rm infr}
                + O(\delta) \ .
\eeq
Here, we have introduced the correction $\Delta E^{N1}_{P,\rm infr}$ that does not
depend on the regulator $\delta$ and can be obtained by fitting numerical
results for $\Delta E^{N1}_{P, IR}(\delta)$.

An analogous evaluation for the IR-divergent part of $\Delta E^{N2}_P$ yields
\beq                                            \label{bf13}
\Delta E^{N2}_{P, IR}(\delta) = \frac{\alpha}{\pi}
        \Delta E^{(1)}_{\rm one} \ln \delta + \Delta E^{N2}_{P,\rm infr}
                + O(\delta) \ .
\eeq
As can be seen, the IR-divergent parts separated in Eqs. (\ref{bf12}) and (\ref{bf13})
are exactly cancelled by the corresponding terms in Eqs. (50) and (55) of I.

%
\section{Numerical evaluation}

\subsection{Green function in the mixed representation} \label{idea}

The main problem of the numerical evaluation of the P terms is that they involve the
Dirac-Coulomb Green function in momentum space. Until recently, there has been no
convenient method for its evaluation. As was pointed out in I, new numerical tools
should be developed for the calculation of the P terms.

Here we address two main features which allow us to evaluate the P terms. Firstly, we
express them in a form which involves the Fourier transform of the Green function over
only one radial variable [Eqs. (\ref{bf2a}) and (\ref{bf2b})]. We refer to this as the
mixed (momentum-coordinate) representation. Secondly, we develop a convenient
scheme for the numerical evaluation of the Green function in the mixed representation.
This scheme was proposed and tested in our previous evaluation of the irreducible part
of the diagram in Fig.~\ref{sese}(a) (the loop-after-loop contribution)
\cite{Yerokhin00}. Here we describe the basic idea of this approach.

We start from the B-splines method for the Dirac equation \cite{Johnson88}. For a
fixed angular-momentum quantum number $\kappa$, it provides a finite set of radial
wave functions
$
\left\{ \varphi^i_{\kappa,n}(x) \right\}_{n = 1}^{N} \ ,
$
where the superscript $i=1,2$ indicates the upper and the lower component
of the radial wave function, and $n$ numerates the
wave functions in the set. The wave
functions are found
as a linear combination of the B-splines \cite{deBoor},
\beq                                            \label{num2}
\varphi^i_{\kappa,n}(x)  = \frac1x \sum_l a^i(\kappa,n,l) B_{l}(x)\ ,
\eeq
where $\left\{ B_{l}(x) \right\}$, $l = 1,2,\ldots$ is the set of the B-splines
defined on the grid $\left\{ x_l \right\}$. Since each of $B_{l}(x)$ can be
represented as a piecewise polynomial, we can build the corresponding
piecewise-polynomial representation for the wave functions,
\begin{equation}                                \label{num3}
 \varphi_{\kappa,n}^i(x) = \frac1{x} \sum_k c^i_k(\kappa,n,l) \,
        (x-x_l)^k \ , \ \ \ \ x\in \left[ x_l,x_{l+1} \right] \ .
\end{equation}

Consequently, the radial Dirac-Coulomb Green function
in the coordinate space, defined as
\begin{equation}                                \label{num4}
  G^{ij}_{\kappa}(\vare,x_1,x_2) = \sum_n
  \frac{\varphi_{\kappa,n}^i(x_1)\varphi_{\kappa,n}^j(x_2)}{\vare-\vare_n} \ ,
\end{equation}
can be written in an analogous form,
\beqn                           \label{num5}
  G_{\kappa}^{ij}(\vare,x_1,x_2) &=& \frac1{x_1x_2} \sum_{k_1k_2}
      A^{ij}_{k_1k_2}(\vare,\kappa,l_1,l_2) (x_1-x_{l_1})^{k_1}
                (x_2-x_{l_2})^{k_2}\ ,
        \nonumber \\ &&
x_1\in \left[ x_{l_1},x_{l_1+1} \right], x_2\in \left[
                x_{l_2},x_{l_2+1} \right] \ .
\eeqn
Here, the coefficients $A^{ij}_{k_1k_2}$ are given by
\begin{equation}                                \label{num6}
    A^{ij}_{k_1k_2}(\vare,\kappa,l_1,l_2) =
    \sum_n \frac{c^i_{k_1}(\kappa,n,l_1) c^j_{k_2}(\kappa,n,l_2)}
        {\vare-\vare_n} \ .
\end{equation}
At this point, we have built the Dirac-Coulomb Green function in the
piecewise-polynomial form. This representation is very convenient for the numerical
evaluation. After the coefficients $A^{ij}_{k_1k_2}$ are stored for given values of
$\kappa$ and $\vare$, the computation of the Green function is reduced to the
evaluation of a simple polynomial over each of the radial variables. We note one
additional advantage of this representation of the Green function, as compared to its
closed analytical form. The Green function in the form (\ref{num5}) and its derivative
are continuous functions of the radial arguments, while its analytic representation
contains a discontinuous function $\theta(x_1-x_2)$ (see, {\it e.g.}, \cite{MohrPl98}).

Now we turn to the Green function in the mixed representation. The Fourier transform
of the radial Green function over the second radial argument can be written as
\beq                                            \label{num7}
  G_{\kappa}^{ij}(\vare,x_1,p_2) = 4\pi s(L_j)
        \int_0^{\infty} dx_2 \, x_2^2 \, j_{L_j}(p_2 x_2) \,
                  G_{\kappa}^{ij}(\vare,x_1,x_2) \ ,
\eeq
where $L_{1,2} = |\kappa\pm 1/2|-1/2$, $s(L_1)=1$,
$s(L_2) = -\kappa/|\kappa|$, and $j_L(z)$ denotes the spherical
Bessel function. Introducing the Fourier-transformed basic
polynomials,
\beq                                            \label{num8}
   \Pi^{ik}_{l}(p) = 4\pi s(L_i) \int_{x_l}^{x_{l+1}} dx\,
       x(x-x_l)^k j_{L_i}(px) \ ,
\eeq
we write the Green function in the mixed representation,
\beq                                            \label{num9}
  G_{\kappa}^{ij}(\vare,x_1,p_2) =
        \frac1{x_1}\sum_{k_1}(x_1-x_{l_1})^{k_1}
           \sum_{l_2 k_2} A^{ij}_{k_1k_2}(\vare,\kappa,l_1,l_2)
                 \Pi^{jk_2}_{l_2}(p_2) \ , \ \ \
                x_1\in \left[ x_{l_1},x_{l_1+1} \right] \ .
\eeq
Certainly, a computation in the mixed representation is essentially more
time-consuming than that in coordinate space, due to necessity to evaluate the whole
set of the integrals $\Pi^{jk_2}_{l_2}(p_2)$ for each new value of $p_2$. Still, in
actual calculations we can perform the numerical integration over $x_1$ first, and the
total amount of computational time can be kept very reasonable.

\subsection{Evaluation of $\Delta E^{N1}_P$}

First, we separate the total contribution $\Delta E^{N1}_P$ into two pieces, the
IR-divergent part $\Delta E^{N1}_{P, IR}$ given by Eq.~(\ref{bf8}), and the finite
remainder $\Delta E^{N1}_{P,r}$. The expression for $\Delta E^{N1}_{P,r}$ is given by
\beqn                                           \label{num10}
\Delta E^{N1}_{P,r} &=&
        \frac{i\alpha}{2\pi} \intinf d\omega
                \int \frac{d\bfp}{(2\pi)^3}
                        \int d\bfx_1 d \bfx_2 \,
        \frac{\exp(i|\omega|x_{12})}{x_{12}}
                \nonumber \\ && \times
                 \psi^{\dag}_a(\bfx_1) \alpha_{\mu}
      {\cal  G}_{N1}(\vare_a-\omega,\bfx_1,\bfp,\bfx_2)
         \alpha^{\mu}        \psi_a(\bfx_2)
\ ,
\eeqn
where
\beqn                                           \label{num11}
{\cal G}_{N1}(\vare,\bfx_1,\bfp,\bfx_2) &=&
        \ \ G(\vare,\bfx_1,\bfp)
                    \gamma_0  \Sigma^{(0)}_R(\vare,\bfp)
                        G(\vare,\bfp,\bfx_2)
                \nonumber \\ &&
        -G^{(0)}(\vare,\bfx_1,\bfp)
                    \gamma_0  \Sigma^{(0)}_R(\vare,\bfp)
                        G^{(0)}(\vare,\bfp,\bfx_2)
                \nonumber \\ &&
        -G^{(1)}(\vare,\bfx_1,\bfp)
                    \gamma_0  \Sigma^{(0)}_R(\vare,\bfp)
                        G^{(0)}(\vare,\bfp,\bfx_2)
                \nonumber \\ &&
        -G^{(0)}(\vare,\bfx_1,\bfp)
                    \gamma_0  \Sigma^{(0)}_R(\vare,\bfp)
                        G^{(1)}(\vare,\bfp,\bfx_2)
                \nonumber \\ &&
        -\sum_{\mu_{\overline{a}}}
          \frac{\psi_{\overline{a}}(\bfx_1)
                   \psi^{\dag}_{\overline{a}}(\bfp)}{\vare-\vare_a+i0}
                    \gamma_0  \Sigma^{(0)}_R(\vare,\bfp)
          \frac{\psi_{\overline{a}}(\bfp)
                   \psi^{\dag}_{\overline{a}}(\bfx_2)}{\vare-\vare_a+i0}
        \ .
\eeqn

The next step is to perform the Wick rotation of the $\omega$-integration contour in
Eq.~(\ref{num10}). This is very convenient for the numerical evaluation since,
firstly, the Dirac Green function as well as the photon propagator are exponentially
decaying functions for imaginary values of $\omega$. Secondly, by this deformation of
the contour we escape most of the problems connected with poles of the electron
propagators and with the analytic structure of $\Sigma^{(0)}_R(\vare)$. The analysis
given in Appendix~\ref{Apse} shows that $\Sigma^{(0)}_R(\vare_a-\omega)$ as a function
of real $\omega$ allows the analytical continuation to the first quadrant of the
complex plane starting from the right half of the real axis, and to the third quadrant
from the left half of the axis. Therefore, we can rotate the integration contour on
the imaginary axis dividing $\Delta E^{N1}_{P, r}$ into two pieces, $\Delta
E^{N1}_{P,\rm Im}$ corresponding to the integral along the imaginary axis, and the
pole term $\Delta E^{N1}_{P,\rm pole}$ that arises from the pole of electron
propagator with $\vare_n = \vare_a$. (At this moment, we assume that $a$ corresponds
to the $1s$-state.) We have
\beqn                                           \label{num13333}
\Delta E^{N1}_{P,\rm Im} &=&
        -\frac{\alpha}{\pi} \Re \int_0^{\infty} d\omega
                \int \frac{d\bfp}{(2\pi)^3}
                        \int d\bfx_1 d \bfx_2 \,
        \frac{\exp(-\omega x_{12})}{x_{12}}
                \nonumber \\ && \times
                 \psi^{\dag}_a(\bfx_1) \alpha_{\mu}
      {\cal  G}_{N1}(\vare_a-i\omega,\bfx_1,\bfp,\bfx_2)
         \alpha^{\mu}        \psi_a(\bfx_2)
\ ,
\eeqn
\beqn                                           \label{num14}
\Delta E^{N1}_{P,\rm pole} &=&
        \frac{\alpha}{2} \Re
                \int \frac{d\bfp}{(2\pi)^3}
                        \int d\bfx_1 d \bfx_2 \,
        \frac{1}{x_{12}}
                 \psi^{\dag}_a(\bfx_1) \alpha_{\mu}
     {\cal  G}_{N1}^{\rm pole}(\vare_a,\bfx_1,\bfp,\bfx_2)
            \alpha^{\mu}        \psi_a(\bfx_2)
\ ,
\eeqn
where
\beqn                                           \label{num15}
{\cal G}_{N1}^{\rm pole}(\vare_a,\bfx_1,\bfp,\bfx_2) &=&
        \sum_{\mu_{\overline{a}}} \Bigl[
          \psi_{\overline{a}}(\bfx_1)
                   \psi^{\dag}_{\overline{a}}(\bfp)
                    \gamma_0  \Sigma^{(0)}_R(\vare_a,\bfp)
                        G^{\rm red}(\vare_a,\bfp,\bfx_2)
                \nonumber \\ &&
         +
        G^{\rm red}(\vare_a,\bfx_1,\bfp)
                    \gamma_0  \Sigma^{(0)}_R(\vare_a,\bfp)
        \psi_{\overline{a}}(\bfp)
                   \psi^{\dag}_{\overline{a}}(\bfx_2)
        \Bigr]
        \ ,
\eeqn
and $G^{\rm red}$ is the reduced Dirac-Coulomb Green function,
\beq                                            \label{num16}
G^{\rm red}(\vare,\bfx_1,\bfx_2) = \sum^{\vare_n \ne \vare_a}_n
        \frac{ \psi_n(\bfx_1) \psi^{\dag}_n(\bfx_2)}
                {\vare-\vare_n(1-i0)} \ .
\eeq

Finally, we collect all contributions to $\Delta E^{N1}_P$,
\beq                                            \label{num17}
\Delta E^{N1}_{P}(\delta) = \frac{\alpha}{\pi}
        \Delta E^{(1)}_{\rm zero} \ln \delta
                 + \Delta E^{N1}_{P,\rm infr}
        + \Delta E^{N1}_{P,\rm pole}+ \Delta E^{N1}_{P,\rm Im}
                + O(\delta) \ .
\eeq
We note that instead of dividing $\Delta E^{N1}_{P}$ into three parts, we could have
introduced the regulator $\delta$ right from the beginning in $\Delta E^{N1}_{P}$, as
it was done in I for the ``M terms''. However, this would cause a rapidly varying
structure of the integrand in the low-$\omega$ region, which makes calculations much
more time-consuming. ({\it E.g.,} for the pole term introducing a regulator would
involve a numerical evaluation of the integral which yields the $\delta$-function in
the limit $\delta \to 0$.) In our approach, on the contrary, only the IR-divergent
part is evaluated with the regulator; the corresponding calculation is relatively
simple and allows accurate fitting to the form (\ref{bf12}).

Let us now outline essential features of our numerical evaluation. As can be seen, the
dependence of the functions ${\cal G}_{N1}$ and ${\cal G}_{N1}^{\rm pole}$ on the
angular parts of $\bfx_1$ and $\bfx_2$ is exactly the same as that of the Dirac Green
function. Therefore, the angular integration causes no problems and can be performed
by a straightforward generalization of formulas given in Ref. \cite{Yerokhin99}. The
most problematic part of the numerical evaluation of $\Delta E^{N1}_{P}$ is the
calculation of $\Delta E^{N1}_{P,\rm Im}$. All numerical integrations were performed
by the Gauss-Legendre quadratures. The ordering of integrations in our computation
coincides with that of Eq.~(\ref{num13333}), with the summation over the
angular-momentum quantum number of intermediate states $\kappa$ moved outside. For
each value of $\kappa$ and $\omega$, we store three sets of complex coefficients
$A^{ij}_{k_1k_2}$ corresponding to the functions $G$, $G^{(0)}$, and $G^{(1)}$. For
each value of $p$, we calculate also a set of the Fourier-transformed polynomials
$\Pi^{ik}_l(p)$. After this, the integrations over the radial variables $x_1$, $x_2$
are performed. This scheme is rather efficient and was used for the evaluation of
$\Delta E^{N1}_{P,\rm Im}$. The corrections $\Delta E^{N1}_{P,\rm infr}$ and $\Delta
E^{N1}_{P,\rm pole}$ were calculated in several different ways, which served as a good
test for our numerical procedure.

The actual calculation was performed with the basis set constructed typically with
50-60 B-splines of order 6 and employing the point nuclear model. The cavity size of 1
a.u. was employed for $Z=80$, which was scaled as $\gamma/Z$ with $Z$, $\gamma =
\sqrt{1-(\alpha Z)^2}$. We use an exponential grid with the first knot of about 0.001
a.u. for $Z=80$. This particular grid was chosen since it yields an optimal
convergence in the evaluation of the first-order self-energy correction with respect
to the number of knots. The infinite summation over the angular-momentum quantum
number of intermediate states $\kappa$ was terminated typically at $|\kappa_{\rm
max}|=7$. The tail of the expansion was estimated by polynomial fitting in
$1/|\kappa|$. The results of the numerical evaluation of the individual contributions
to $\Delta E^{N1}_P$ are presented in Table \ref{Table_dEN1} for $Z=83$, $90$, and
$92$. The numerical errors, quoted in the table, originate mainly from the sensitivity
of the result to the number of knots and different grids.

\subsection{Evaluation of $\Delta E^{N2}_P$}

The correction $\Delta E^{N2}_P$ can be written in the same
way as $\Delta E^{N1}_P$,
\beq                                            \label{num18}
\Delta E^{N2}_{P}(\delta) = \frac{\alpha}{\pi}
        \Delta E^{(1)}_{\rm one} \ln \delta
                 + \Delta E^{N2}_{P,\rm infr}
        + \Delta E^{N2}_{P,\rm pole}+ \Delta E^{N2}_{P,\rm Im}
                + O(\delta) \ .
\eeq
Here, we again separate the IR-divergent part [Eq.~(\ref{bf13})] and perform the Wick
rotation of the $\omega$-integration contour, separating the corresponding pole
contribution ($\Delta E^{N2}_{P,\rm pole}$). We note that the rotation of the contour
is possible because the vertex operator $\Gamma^{\mu}(\vare_a-\omega,\vare_a-\omega)$
as a function of real $\omega$ allows an analytic continuation in the region of
interest, as shown in Appendix~\ref{Apver}. The resulting expression reads
\beqn                                           \label{num19}
\Delta E^{N2}_{P,\rm Im} &=&
        -\frac{\alpha}{\pi} \Re \int_0^{\infty} d\omega
                \int \frac{d\bfp_1}{(2\pi)^3}
                 \frac{d\bfp_2}{(2\pi)^3}
                        \int d\bfx_1 d \bfx_2 \,
        \frac{\exp(-\omega x_{12})}{x_{12}}
      V_C(\bfq)
                \nonumber \\ && \times
                 \psi^{\dag}_a(\bfx_1) \alpha_{\mu}
        {\cal  G}_{N2}(\vare_a-i\omega,\bfx_1,\bfp_1,\bfp_2,\bfx_2)
         \alpha^{\mu}        \psi_a(\bfx_2)
\ ,
\eeqn
where
\beqn                                           \label{num20}
{\cal G}_{N2}(\vare,\bfx_1,\bfp_1,\bfp_2,\bfx_2) &=&
        \ \ G(\vare,\bfx_1,\bfp_1)
                \gamma_0  \Gamma^0_R(\vare,\bfp_1;\vare,\bfp_2)
                        G(\vare,\bfp_2,\bfx_2)
                \nonumber \\ &&
         -G^{(0)}(\vare,\bfx_1,\bfp_1)
                \gamma_0  \Gamma^0_R(\vare,\bfp_1;\vare,\bfp_2)
                        G^{(0)}(\vare,\bfp_2,\bfx_2)
                \nonumber \\ &&
        -\sum_{\mu_{\overline{a}}}
          \frac{\psi_{\overline{a}}(\bfx_1)
                   \psi^{\dag}_{\overline{a}}(\bfp_1)}{\vare-\vare_a+i0}
                \gamma_0  \Gamma^0_R(\vare,\bfp_1;\vare,\bfp_2)
          \frac{\psi_{\overline{a}}(\bfp_2)
                   \psi^{\dag}_{\overline{a}}(\bfx_2)}{\vare-\vare_a+i0}
        \ .
\eeqn
Assuming that $a$ is the $1s$ state, the pole contribution is given by
\beqn                                           \label{num21}
\Delta E^{N2}_{P,\rm pole} &=&
         \frac{\alpha}{2} \Re
                \int \frac{d\bfp_1}{(2\pi)^3}
                 \frac{d\bfp_2}{(2\pi)^3}
                        \int d\bfx_1 d \bfx_2 \,
        \frac{1}{x_{12}} V_C(\bfq)
                \nonumber \\ && \times
       \psi^{\dag}_a(\bfx_1) \alpha_{\mu}
       {\cal  G}_{N2}^{\rm pole}(\vare_a,\bfx_1,\bfp_1,\bfp_2,\bfx_2)
         \alpha^{\mu}        \psi_a(\bfx_2)
\ ,
\eeqn
where
\beqn                                           \label{num22}
{\cal G}_{N2}^{\rm pole}(\vare_a,\bfx_1,\bfp_1,\bfp_2,\bfx_2) &=&
        \sum_{\mu_{\overline{a}}} \Bigl[
          \psi_{\overline{a}}(\bfx_1)
                   \psi^{\dag}_{\overline{a}}(\bfp_1)
                \gamma_0  \Gamma^0_R(\vare_a,\bfp_1;\vare_a,\bfp_2)
                        G^{\rm red}(\vare_a,\bfp_2,\bfx_2)
                \nonumber \\ &&
         +
        G^{\rm red}(\vare_a,\bfx_1,\bfp_1)
                \gamma_0  \Gamma^0_R(\vare_a,\bfp_1;\vare_a,\bfp_2)
        \psi_{\overline{a}}(\bfp_2)
                   \psi^{\dag}_{\overline{a}}(\bfx_2)
        \Bigr]
        \ .
\eeqn

The angular integration in these expressions is straightforward, due to the fact that
the functions ${\cal G}_{N2}$ and ${\cal G}_{N2}^{\rm pole}$ have the same dependence
on the angular parts of $\bfx_1$ and $\bfx_2$ as the Dirac Green function $G$. The
numerical evaluation of the correction $\Delta E^{N2}_{P,\rm Im}$ is much more
time-consuming than that of $\Delta E^{N1}_{P,\rm Im}$. This is because the
integration over $|\bfp|$ in Eq.~(\ref{num13333}) (after the angular integration has
been carried out) is substituted by the triple integration over $|\bfp_1|$,
$|\bfp_2|$, and $\xi = (\hat{\bfp}_1\cdot\hat{\bfp}_2)$. So, the numerical evaluation
of $\Delta E^{N2}_{P,\rm Im}$ involves one infinite partial-wave summation and a
seven-fold numerical integration. (One additional integral comes from the evaluation
of the Green function in the mixed representation.) While it is possible to evaluate
$\Delta E^{N2}_{P,\rm Im}$ in a similar way as $\Delta E^{N1}_{P,\rm Im}$, we have
found a more efficient method for its computation. To this end, we rewrite
Eq.~(\ref{num19}) as follows
\beqn                                           \label{num23}
\Delta E^{N2}_{P,\rm Im} &=&
        -\frac{\alpha}{\pi} \Re \int_0^{\infty} d\omega
                \int \frac{d\bfp_1}{(2\pi)^3}
                \frac{d\bfp_2}{(2\pi)^3} \,
      V_C(\bfq)
                \nonumber \\ && \times
        \Biggl\{ \sum_{n_1n_2}{}\pr
          \frac{ \psi^{\dag}_{n_1}(\bfp_1)\, \gamma_0 \Gamma^0_R\, \psi_{n_2}(\bfp_2)}
                {(\vare_a-i\omega-\vare_{n_1})(\vare_a-i\omega-\vare_{n_2})}
        \lbr an_2|
                \frac{1-\balpha_1\balpha_2}{x_{12}}
                        e^{-\omega x_{12}}|n_1a\rbr
                \nonumber \\ &&
- \sum_{{\beta}{\gamma}} \frac{u^{\dag}_{{\beta}}(\bfp_1)\,  \gamma_0
            \Gamma^0_R\, u_{{\gamma}}(\bfp_2)}
    {(\vare_a-i\omega-\vare_{{\beta}})(\vare_a-i\omega-\vare_{{\gamma}})}
\lbr a{\gamma}| \frac{1-\balpha_1\balpha_2}{x_{12}}
                        e^{-\omega x_{12}}|{\beta}a\rbr
                        \Biggr\} \ ,
\eeqn
where $\Gamma^0_R \equiv \Gamma^0_R(\vare_a-i\omega,\bfp_1;\vare_a-i\omega,\bfp_2)$,
$\psi_{n_1}$ and $\psi_{n_2}$ stand for solutions of the Dirac equation with the
Coulomb potential, $u_{\beta}$ and $u_{\gamma}$ denote solutions of the free Dirac
equation, and the prime on the sum indicates that the term with $\vare_{n_1} =
\vare_{n_2} = \vare_a$ is omitted. In order to evaluate Eq.~(\ref{num23}), we
introduce the matrix $S$,
\beq                                            \label{num24}
S^{ij}_{\kappa}(\omega,p_1,p_2) =
        \sum_{n_1n_2}{}\pr
          \frac{\varphi^i_{\kappa,n_1}(p_1) \varphi^j_{\kappa,n_2}(p_2)}
        {(\vare_a-i\omega-\vare_{n_1})
                (\vare_a-i\omega-\vare_{n_2})}
\lbr an_2|
                \frac{1-\balpha_1\balpha_2}{x_{12}}
                        e^{-\omega x_{12}}|n_1a\rbr     \ ,
\eeq
where $\varphi^i(p)$ ($i=1,2$) stands for the radial components of the corresponding
wave function. An analogous matrix $S^{(0)}$ is introduced also for the second part of
Eq.~(\ref{num23}). For each value of $\kappa$ and $\omega$, we calculate coefficients
of the piecewise-polynomial representation of $S$, $S^{(0)}$. Then, for each values of
$p_1$ and $p_2$, we store two sets of the Fourier-transformed polynomials,
$\Pi^{ik}_l(p_1)$ and $\Pi^{ik}_l(p_2)$. Finally, the integration over $\xi$ is
performed.

The results of the numerical evaluation
of the individual contributions to $\Delta E^{N2}_P$ are presented in Table
\ref{Table_dEN2} for $Z=83$, $90$, and $92$.

\subsection{Evaluation of $\Delta E^{O1}_P$}

The expression for $\Delta E^{O1}_P$ can be easily obtained from Eq.~(\ref{bf4}),
after rewriting it in terms of the Green function and making the substitutions $G \to
G^{(2+)}$ and $\Gamma^{\mu} \to \Gamma^{\mu}_R$,
\beqn                                           \label{num25}
\Delta E^{O1}_P &=&
-2i\alpha \intinf d\omega \int \frac{d\bfp_1}{(2\pi)^3}
        \frac{d\bfp_2}{(2\pi)^3} \int d\bfz \,
                \frac{\exp(-i\bfq \bfz)}{\omega^2 -\bfq^2+ i0}
                \nonumber \\ && \times
\psi_a^{\dag}(\bfz) \alpha_{\mu} G^{(2+)}(\vare_a-\omega,\bfz,\bfp_1)
        \gamma_0 \Gamma^{\mu}_R(\vare_a-\omega,\bfp_1;\vare_a,\bfp_2)
                        \psi_a(\bfp_2)
\ ,
\eeqn
where $\bfq = \bfp_1-\bfp_2$, $G^{(2+)}(\vare,\bfz,\bfp_1)$ is the many-potential
Green function [Eq.~(\ref{eq9})] in the momentum-coordinate representation. The
analysis given in Appendix~\ref{Apver} shows that the vertex operator
$\Gamma^{\mu}_R(\vare_a-\omega,\vare_a)$ as a function of real $\omega$ allows the
analytical continuation to the first quadrant of the complex $\omega$ plane starting
from the right half of the real axis, and to the third quadrant from the left half of
the axis. Therefore, we can perform the Wick rotation of the integration contour,
separating the corresponding pole contribution,
\beq                                            \label{num26}
\Delta E^{O1}_P = \Delta E^{O1}_{P,\rm pole} + \Delta E^{O1}_{P,\rm Im}
 \ .
\eeq

The angular integration in Eq.~(\ref{num25}) is by far more difficult as compared to
the contributions considered so far. As the involved expressions are rather lengthy,
we do not give their detailed consideration here. However, in order to give the reader
an idea how the angular integration is performed, we note that Eq.~(\ref{bf4}) is very
similar to the free-vertex contribution which is encountered in a calculation of the
self-energy screening diagrams (compare with Eq.~(68) in Ref. \cite{Yerokhin99a}).
Basically, the angular integration in Eq.~(\ref{num25}) is the same as described in
detail in Ref. \cite{Yerokhin99a}. The only difference is that in Ref.
\cite{Yerokhin99a} the integration was demonstrated for two particular states,
$n=2p_{1/2}$ and $2s$. In Eq. (\ref{num25}) we need a generalization of that procedure
for an arbitrary $n$, which is somewhat tedious but straightforward.

Let us discuss now the numerical evaluation of $\Delta E^{O1}_P$.
After the angular integration is carried out, a typical contribution
to $\Delta E^{O1}_{P,\rm Im}$ can be written as follows
\beqn                                           \label{num27}
t &=& \sum_{\kappa} \int_0^{\infty} d\omega  \int_0^{\infty} dp_1
        \int_0^{\infty} dp_2 \int_{-1}^1 d \xi \int_0^{\infty} dz \,
                \frac{p_1^2\, p_2^2\, z^2}{\omega^2 + q^2}
                \nonumber \\ && \times
j_l(q z) \varphi^i_a(z) G_{\kappa}^{(2+)^{ij}}(\vare_a-i\omega,z,p_1)
        f(p_1,p_2,\xi) \varphi_a^j(p_2)  \ ,
\eeqn
where $p_1 = |\bfp_1|$, $p_2 = |\bfp_2|$, $\xi = (\hat{\bfp}_1\cdot\hat{\bfp}_2)$,
$q^2 = p_1^2+p_2^2-2p_1p_2\xi$, $\varphi^i$ is a radial component of the wave
function, $G^{(2+)^{ij}}_{\kappa}$ denotes a radial component of $G^{(2+)}$, $j_L$ is
a spherical Bessel function, and $f(p_1,p_2,\xi)$ originates from the vertex operator.
Eq.~(\ref{num27}) involves one infinite partial-wave summation and a six-fold
numerical integration. (The sixth integral is the momentum integration in the
evaluation of the Green function in the mixed representation. Two additional
integrations can be also mentioned, one over the Feynman parameter in the evaluation
of $f(p_1,p_2,\xi)$, and another a radial integration in the computation of $G^{(1)}$.
This makes the total dimension of the integral to be 8.) Two of these integrations
involve spherical Bessel functions, which oscillate strongly in the high-momenta
region. In order to keep the amount of computational time within an acceptable limit,
the general scheme of the calculation should be chosen carefully. In our approach, we
introduce the following change of variables \cite{Yerokhin99}: $\left\{ p_1,p_2,\xi
\right\} \to \left\{ x,y,q \right\}$, where $x = p_1+p_2$, $y = p_>-p_<$, $q^2 =
p_1^2+p_2^2-2p_1p_2\xi$, $p_> = \max(p_1,p_2)$, $p_< = \min(p_1,p_2)$. After that, we
have
\beq                            \label{num28}
\int_0^{\infty} dp_1 \int_0^{\infty} dp_2 \int_{-1}^1 d\xi \, F(p_1,p_2,\xi) =
\int_0^{\infty} dx \int_0^x dy \int_y^x d q  \, \frac{q}{2p_1p_2} \Bigl[
F(p_1,p_2,\xi) + F(p_2,p_1,\xi) \Bigr] \ .
\eeq

In the actual calculation, the outermost loop was the summation over $\kappa$. The
next loop is the $\omega$ integration. For given values of $\kappa$ and $\omega$, we
store a set of coefficients of the piecewise-polynomial representation of
$G^{(2+)}_{\kappa}$. These coefficients are obtained as the difference of the
corresponding coefficients for $G_{\kappa}$, $G_{\kappa}^{(0)}$, and
$G^{(1)}_{\kappa}$, as stated in Eq. (\ref{eq10}). The next two loops are the
integrations over $x$ and $y$. For each value of $p_1$, we store a set of the
Fourier-transformed polynomials $\Pi^{ik}_l(p_1)$. The next step is the integration
over $q$ and the evaluation of $f(p_1,p_2,\xi)$, which involves an integration over
the Feynman parameter (see, {\it e.g.}, Ref. \cite{Yerokhin99}). The innermost loop is
the integration over $z$. Its optimization is the most critical part from the point of
view of computational time. For small values of $q$, we use Gauss-Legendre
quadratures. When $q$ is large, we decompose $j_l(qx)$ in a combination of $\sin(qx)$
and $\cos(qx)$, and use the standard routine for the $\sin$- and $\cos$-Fourier
transforms based on the generalized Clenshaw-Curtis algorithm. At that stage of the
computation, both $\varphi^i_a$ and $G^{(2+)^{ij}}_{\kappa}$ are represented by
piecewise polynomials and, therefore, the $z$ integration can be performed rather
fast.

A good test for our numerical procedure is to evaluate the many-potential part of the
first-oder self-energy correction, which can be obtained from Eq.~(\ref{num25}) by the
substitution $\Gamma^{\mu}_R \to \gamma^{\mu}$.

The results of the numerical calculation of the individual contributions to $\Delta
E^{O1}_P$ are presented in Table \ref{Table_dEO1} for $Z=83$, $90$, and $92$.

%
\section{Results and discussion}

In this paper we present the numerical evaluation of the correction
$\Delta E$, given by the three sets of diagrams which are shown in Figs.
\ref{dEN1}--\ref{dEO1}. Putting together Eqs. (\ref{bf1}), (\ref{bf7}),
(\ref{num17}), (\ref{num18}), and (\ref{num26}), we have
\beqn                                   \label{res1}
\Delta E &=& \tilde{L}^{(1)} \Delta E^{(1)}_{{\rm many},D}
+ \frac{\alpha}{\pi} \biggl[ \Delta E^{(1)}_{\rm zero} +
            \Delta E^{(1)}_{\rm one} \biggr] \ln \delta
        \nonumber \\ &&
+  \Delta E^{N1}_{P,\rm infr}+
        \Delta E^{N1}_{P,\rm pole}+ \Delta E^{N1}_{P,\rm Im}
        \nonumber \\ &&
+ \Delta E^{N2}_{P,\rm infr}
        + \Delta E^{N2}_{P,\rm pole}+ \Delta E^{N2}_{P,\rm Im}
        \nonumber \\ &&
+ 2\Delta E^{O1}_{P,\rm pole} + 2\Delta E^{O1}_{P,\rm Im} + O(\delta) \ .
\eeqn
The UV-finite difference $\Delta E - \tilde{L}^{(1)} \Delta E^{(1)}_{{\rm many},D}$
corresponds to what in I is called the P term. The IR-divergent contributions, still
presented in the P term, are cancelled when considered together with Eqs. (50) and
(55) of I. When all $\delta$-dependent terms are put together, the limit $\delta \to
0$ can be taken, and contributions of order $\delta$ and higher vanish. Finite
individual contributions to $\Delta E^{N1}_P$, $\Delta E^{N2}_P$, and $\Delta
E^{O1}_P$ are listed in Tables \ref{Table_dEN1}--\ref{Table_dEO1}, respectively. In
Table~\ref{Table_dE} we collect all finite contributions to $\Delta E$.

\begin{figure}
\centerline{ \mbox{ \epsfxsize=\textwidth \epsffile{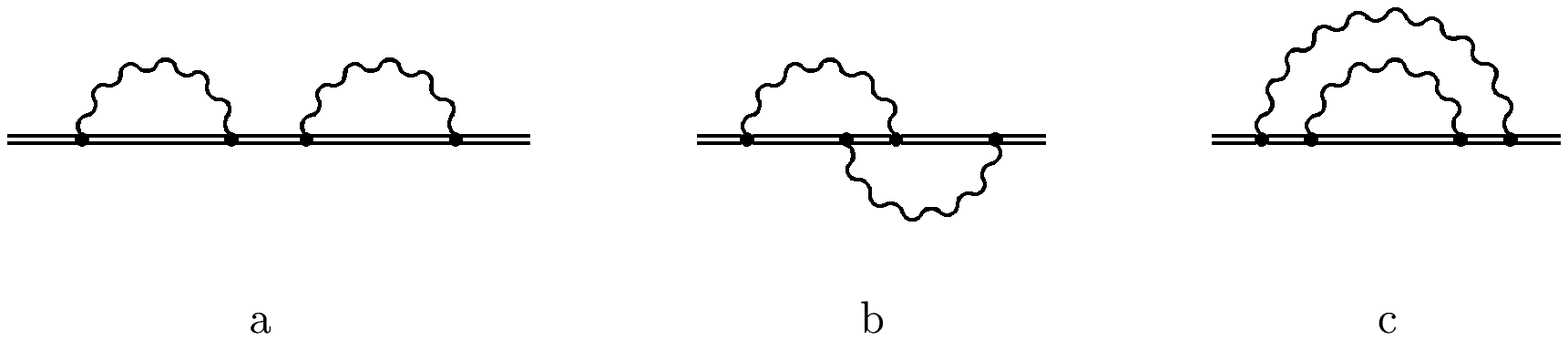} } }
\caption{One-electron self-energy Feynman diagrams of second order in
$\alpha$. \label{sese}}
\end{figure}

Now we can obtain a finite, gauge-independent (within the covariant gauges) result for
the sum of the diagrams in Figs.~\ref{sese}(b,c) and the reducible part of the diagram
in Fig.~\ref{sese}(a). In order to get this, we should add together the contributions
listed in Table~\ref{Table_dE}, the results from I for the M terms (Eqs. (50), (52),
and (55) of I) and those for the F terms (Table IV of I). This yields $-$0.903(11) eV
for the ground state of H-like uranium and $-$0.575(11) eV for bismuth. Adding this to
the irreducible part of the diagram in Fig.~\ref{sese}(a) ($-$0.971 eV for $Z=92$
\cite{Mitrushenkov95,Mallampalli98a} and $-$0.544 eV for $Z=83$, this work), we have
for the total two-loop self-energy correction, given by the diagrams in
Fig.~\ref{sese}, $-$1.874(11) eV for $Z=92$ and $-$1.119(11) eV for $Z=83$.

The numerical results for the sum of the diagrams in Figs.~\ref{sese}(b,c) and the
reducible part of the diagram in Fig.~\ref{sese}(a), obtained by combining the present
calculation with that of I, can be compared with the recent evaluation announced in
Ref. \cite{Goidenko01}. The results of $-$0.903(11) eV ($Z=92$) and $-$0.575(11) eV
($Z=83$) obtained in this work should be compared with 1.28(15) eV and 0.73(9) eV,
respectively, reported in Ref. \cite{Goidenko01}. Surprisingly enough, the comparison
shows that the results disagree even with respect to the overall sign of the
contribution. Commenting this disagreement, one can mention that the partial-wave
renormalization procedure, used in Ref.~\cite{Goidenko01}, is known to produce certain
spurious terms due to the noncovariant nature of the regularization (see Ref.
\cite{PerssonPWR} and a discussion given in Refs.~\cite{Yerokhin01hydr,Yerokhin00}).
We also note that some assumptions employed in the numerical evaluation of
Ref.~\cite{Goidenko01} make it difficult to keep accuracy under proper control
in the computation. Still, in order to resolve this disagreement, it is desirable to
perform an evaluation of the total two-loop self-energy correction within the
covariant approach from the beginning up to the end by the same authors. This will be
the aim of our future investigation.

\begin{figure}
\centerline{ \mbox{
\epsfxsize=0.7\textwidth \epsffile{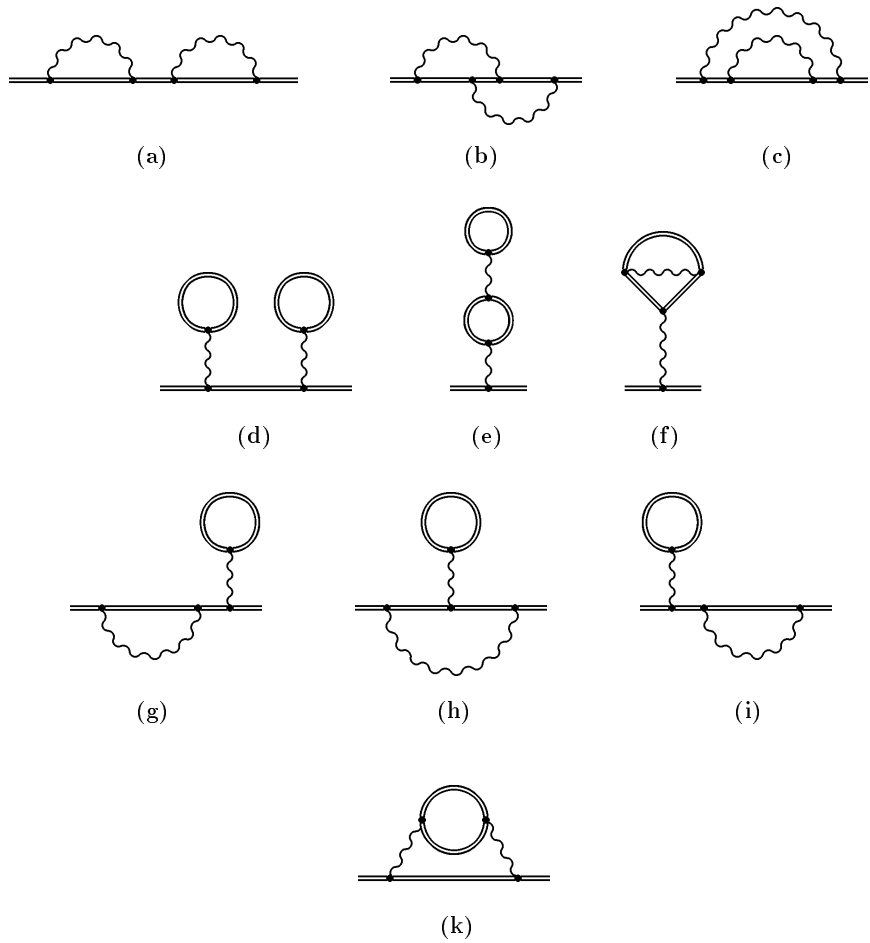}
}}
\caption{One-electron QED corrections of second order in $\alpha$. \label{2order}}
\end{figure}

With this evaluation of the two-loop self-energy correction, we complete the
long-lasting problem of calculation of all second-order (in $\alpha$) QED corrections
for the hydrogen-like ions without an expansion in the parameter $Z\alpha$. The
complete set of these corrections is presented in Fig.~\ref{2order}. The whole set is
conveniently divided into several gauge invariant subsets: SESE (a-c), VPVP (d), VPVP
(e), VPVP (f), SEVP (g-i), S(VP)E (k). In Table~\ref{uranium} we collect all available
contributions to the ground-state Lamb shift in ${}^{238}{\rm U}^{91+}$. The
nuclear-size correction is calculated for the Fermi nuclear model with $\lbr
r^2\rbr^{1/2} = 5.860(2)$ fm \cite{Zumbro84}. The uncertainty of 0.38 eV ascribed to
the nuclear-size effect is evaluated as the difference between the corrections obtained
within the Fermi model and with the homogeneously-charged sphere distribution,
employing the same rms radius. Some of the $\alpha^2$ QED corrections [VPVP (f) and
S(VP)E (k)] are evaluated only within the Uehling approximation at present. We ascribe
the uncertainty of 50\% to them. To obtain the binding energy, the Dirac point-nucleus
eigenvalue of $-$132279.92(1) eV should be added to the Lamb shift presented in
Table~\ref{uranium}. The error of 0.01 eV of the Dirac binding energy results from the
uncertainty of the Rydberg constant \cite{Mohr00}. As can be seen from the table, the
present level of experimental precision provides a test of QED effects of first order
in $\alpha$ on the level of 5\%.

With the two-loop self-energy calculated for the $1s$ state, we can estimate its value
for the $2p_{1/2}$-$2s$ transition in Li-like ions. As is known, the leading
contribution to the self-energy arises from small distances, where the self-energy
operator is close to a $\delta$-functional potential. This gives the well-known
$1/n^3$ scaling for the $s$-states and zero for the $p$-states. Assuming this scaling,
we have a 0.23(20) eV contribution for the one-electron two-loop self-energy
correction to the $2p_{1/2}$-$2s$ transition energy in Li-like uranium. Now we collect
all second-order QED contributions to this transition energy in uranium, as shown in
Table~\ref{Liuranium}. In the first line of the table, our previous result of
280.48(11) eV \cite{Yerokhin01} is given, in which all available contributions are
included, except one-electron second-order QED effects. The total value of the
transition energy amounts to 280.64(11)(21) eV, which agrees well with the
experimental result of 280.59(10) eV \cite{Schweppe91}. In the theoretical prediction,
the first quoted error arises from the uncertainty due to the finite nuclear size
effect and due to higher-order electron correlations (see discussion in Ref.
\cite{Yerokhin01}). The second quoted error corresponds to the uncertainty of the
second-order one-electron QED effects.

%
\section{Conclusion}

In this paper we developed a convenient numerical approach to the evaluation of
two-loop corrections in the mixed momentum-coordinate representation. The elaborated
scheme was applied to the evaluation of a part of the two-loop self-energy correction
which was omitted in the previous study \cite{Mallampalli98}. It is worth mentioning
that our numerical procedure is relatively cheap from computational point of view, as
compared to \cite{Mallampalli98}. While in that work the total computational time of
about 7000 h was reported for a given value of $Z$, our evaluation requires only
100-150 h on a Pentium-like computer for one value of $Z$.

The results of our calculation combined with those of \cite{Mallampalli98} yield the
total value for the two-loop self-energy correction of $-$1.874(11) eV for the ground
state of H-like uranium and of $-$1.119(11) eV for bismuth. As this correction has been
the last uncalculated second-order QED contribution in these systems up to now, our
calculation improves significantly the accuracy of the theoretical prediction in
one-electron ions. The total result for the ground-state Lamb shift in H-like uranium
amounts to 463.93(50) eV. While the present experimental precision of $\pm$13 eV
\cite{Stoehlker00} is not high enough to test the second-order QED effects, it is
going to be improved by an order of magnitude in the near future \cite{Stoehlker00}.

The evaluation of the two-loop self-energy correction for the $1s$ state allows us to
make an estimate of this contribution for the $2p_{1/2}$-$2s$ transition energy in
Li-like ions. This is of particular importance, since in that case the experimental
accuracy is much better than for H-like ions, which makes Li-like ions very
promising for testing second-order QED effects. The first estimate of the two-loop
self-energy correction allows us to ascribe a well-defined uncertainty to the
theoretical prediction. For the $2p_{1/2}$-$2s$ transition in Li-like uranium, the
total result amounts to 280.64(24) eV, which should be compared with the experimental
value of 280.59(10) eV \cite{Schweppe91}.

%
\section*{Acknowledgement}

VY wishes to thank the Technische Universit\"at Dresden and the Max-Planck Institut
f\"ur Physik Komplexerer Systeme for the hospitality during his visit in 2001. We are
grateful to Natalia Lentsman for improving the language of the paper. This work was supported in
part by the Russian Foundation for Basic Research (Grant No. 01-02-17248) and by the
program "Russian Universities: Basic Research" (project No. 3930).

%
\appendix

\section{One-loop self-energy and vertex operators}
\label{ap1}

The free  one-loop self-energy operator in the Feynman gauge is defined by
\begin{equation}                                \label{ap1a}
\Sigma^{(0)}(\rp) = -4\pi i\alpha
\int \frac{d^D\rk}{(2\pi)^D} \frac1{\rk^2+i0}
\gamma_{\sigma}
\frac{\cross{\rp}-\cross{\rk}+m}{(\rp-\rk)^2-m^2+i0} \gamma^{\sigma} \ .
\end{equation}
To separate UV divergences, we write the self-energy operator in $D = 4-\epsilon$
dimensions as
\beq                                            \label{ap2}
\Sigma^{(0)}(\rp) = \delta m+ \tilde{B}^{(1)} (\cross{\rp}-m)
                        + \Sigma^{(0)}_R(\rp) \ .
\eeq
Here, $\delta m$ is the mass counterterm,
\beq                                            \label{ap3}
\delta m = \frac{3\alpha}{4\pi}m(\Delta_{\epsilon}+ \frac43)
        + O(\epsilon) \ ,
\eeq
$\tilde{B}^{(1)}$ is the UV divergent part of the renormalization constant
$1-Z_2^{-1}$,
\beq                                            \label{ap4}
\tilde{B}^{(1)} = -\frac{\alpha}{4\pi} \Delta_{\epsilon}
                + O(\epsilon)\ ,
\eeq
$\Delta_{\epsilon} = 2/\epsilon -\gamma_E+ \ln 4\pi$, and $\gamma_E$ is the Euler
constant. The contribution $\Sigma^{(0)}_R(\rp)$ is finite; its definition agrees with
that of I [denoted in I as  $\Sigma^{(2:0P)}_c(\rp)$]. We note that, evaluating
two-loop corrections, one should keep terms of order $\epsilon$, since they can yield
a finite contribution when multiplied by a divergent constant of order $1/\epsilon$.
However, in our present evaluation of the P terms we do not need explicit expressions
for these contributions.

The one-loop free vertex operator in the Feynman gauge is given by
\begin{equation}                                \label{ap5}
\Gamma^{\mu}(\rp_1,\rp_2) = -4\pi i \alpha
\int \frac{d^D\rk}{(2\pi)^D}
\frac1{\rk^2+i0}
\gamma_{\sigma}
\frac{\cross{\rp}_1-\cross{\rk}+m}{(\rp_1-\rk)^2-m^2+i0} \gamma^{\mu}
\frac{\cross{\rp}_2-\cross{\rk}+m}{(\rp_2-\rk)^2-m^2+i0} \gamma^{\sigma} \ .
\end{equation}
We define a finite part of the vertex operator through
\beq                                            \label{ap6}
\Gamma^{\mu}(\rp_1,\rp_2) = \tilde{L}^{(1)} \gamma^{\mu}
                        + \Gamma^{\mu}_R(\rp_1,\rp_2) \ ,
\eeq
where $\tilde{L}^{(1)}$ is the UV divergent part of the renormalization constant
$Z_1^{-1}-1$,
\beq                                            \label{ap7}
\tilde{L}^{(1)} = \frac{\alpha}{4\pi} \Delta_{\epsilon}
        + O(\epsilon)\ ,
\eeq
and the Ward identity $\tilde{B}^{(1)} = -\tilde{L}^{(1)}$ is
satisfied. Again, our definition of $\Gamma^{\mu}_R$ exactly
corresponds to that of I ($\Lambda^{(2)}_{c\mu}$ in notations of I).

The explicit expressions for the operators $\Sigma_R^{(0)}$, $\Gamma^{\mu}_R$ in the
limit $\epsilon \to 0$ can be found in Ref. \cite{Yerokhin99}. For their exact
$\epsilon$-dependent form we refer the reader to I.

\section{Analytic properties of the one-loop self-energy operator}
\label{Apse}

In this section we consider analytical properties of the one-loop self-energy operator
$\Sigma^{(0)}(\vare_a-\omega,\bfp)$ as a function of $\omega$. From the definition
(\ref{ap1a}), one can deduce that the self-energy operator can be represented by a
combination of two basic integrals,
\beq                                            \label{si1}
\left\{ J, J_{\mu} \right\} =
        \frac{16\pi^2}{i}  \int \frac{d^D\rk}{(2\pi)^D}
\frac{\left\{1, \rk_{\mu}\right\} } {(\rk^2+i0) [(\rp-\rk)^2-m^2+i0]} \ .
\eeq
By introducing the Feynman parametrization of the denominator, we rewrite this as
\beq                                            \label{si2}
\left\{ J, J_{\mu} \right\} =
        \frac{16\pi^2}{i} \int_0^1 dx
         \int \frac{d^D\rk}{(2\pi)^D}
\frac{\left\{1, \rk_{\mu}\right\} }
        {\bigl[ (\rk-(1-x)\rp)^2- (1-x)(\tilde{m}^2-x\rp^2)\bigr]^2} \ ,
\eeq
where $\tilde{m}^2 = m^2-i0$. Shifting the integration variable $\rk \to
\rk-(1-x)\rp$, we obtain
\beq                                            \label{si3}
\left\{ J, J_{\mu} \right\} =
        \frac{16\pi^2}{i} \int_0^1 dx
        \int  \frac{d^D\rk}{(2\pi)^D}
\frac{\left\{1, (1-x)\rp_{\mu}\right\} }
        {\bigl[ \rk^2- (1-x)(\tilde{m}^2-x\rp^2)\bigr]^2} \ ,
\eeq
where we have taken into account the identity
\beq                                            \label{si3a}
\int  \frac{d^D\rk}{(2\pi)^D} \frac{\rk_{\mu}}{A(\rk^2)} = 0 \ ,
\eeq
with $A(\rk^2)$ being a function of $\rk^2$. The $\rk$ integration yields
\beq                                            \label{si4}
\left\{ J, J_{\mu} \right\} = \frac2{\epsilon}  (4\pi)^{\epsilon/2}
\Gamma(1+\epsilon/2) \int_0^1 dx \frac{\left\{1, (1-x)\rp_{\mu}\right\} }
        {N^{\epsilon/2}}
        \ ,
\eeq
where $N = (1-x)(\tilde{m}^2-x\rp^2)$. In the limit $\epsilon \to 0$, we have
\beq                                            \label{si5}
\frac2{\epsilon} N^{-\epsilon/2} = \frac2{\epsilon}- \ln N + O(\epsilon) \ .
\eeq
Roots of $N$ can be easily found,
\beq                                            \label{si6}
N = x(1-x)(\omega_+-\omega-i0)(\omega_-+\omega-i0) \ ,
\eeq
where $\omega_{\pm} = \sqrt{m^2/x +\bfp^2} \pm \vare_a $ . Obviously, $\omega_+ \ge
\omega_+^0 = m+\vare_a$ and $\omega_- \ge \omega_-^0 = m-\vare_a$ for all values of
$|\bfp|$ and $x \in [0,1]$.

We find that for $\omega \in ]-\omega_-^0,\omega_+^0[$, the integrals $J$, $J_{\mu}$
and, therefore, the self-energy operator $\Sigma^{(0)}(\vare_a-\omega)$ as functions
of real $\omega$ can be analytically continued both into the upper and into the lower
half-plane. However, for $\omega > \omega_+^0$ the self-energy operator allows the
analytical continuation in the upper half-plane only, and for $\omega < -\omega_-^0$
in the lower half-plane only. We can conclude that $\Sigma^{(0)}(\vare_a-\omega)$ is
an analytic function of $\omega$ in the complex plane with the branch cuts
$[m+\vare_a-i0, \infty-i0)$ and $[-m+\vare_a+i0, -\infty+i0)$.

\section{Analytic properties of the vertex operator}
\label{Apver}

Here we are interested in analytical properties of the vertex operator
$\Gamma^{\sigma}(\rp_1,\rp_2)$ as a function of  $\omega$ in two kinematics: a) $\rp_1
= (\vare_a-\omega,\bfp_1)$, $\rp_2 = (\vare_a-\omega,\bfp_2)$; and  b) $\rp_1 =
(\vare_a-\omega,\bfp_1)$, $\rp_2 = (\vare_a,\bfp_2)$.

From the definition (\ref{ap5}), we can deduce that the vertex operator can be written
as a combination of three basic integrals,
\beq                                            \label{ap8}
\left\{ I, I_{\mu}, I_{\mu \nu} \right\} =
       \frac{16\pi^2}{i} \int \frac{d^D\rk}{(2\pi)^D}
\frac{\left\{1;\,  \rk_{\mu};\,  \rk_{\mu} \rk_{\nu} \right\}} {(\rk^2+i0)
[(\rp_1-\rk)^2-m^2+i0][(\rp_2-\rk)^2-m^2+i0]} \ .
\eeq
Introducing the Feynman parametrization of the denominator and shifting the
integration variable $\rk \to \rk-y\rqq-x\rp_2$, we obtain
\beq                                            \label{ap9}
\left\{ I, I_{\mu}, I_{\mu \nu} \right\} =
       \frac{32\pi^2}{i} \int_0^1 dx \int_0^x dy \int \frac{d^D\rk}{(2\pi)^D}
\frac{\left\{1;\,  (y \rqq+x\rp_2)_{\mu};\,  (y\rqq+x\rp_2)_{\mu} (y\rqq+x\rp_2)_{\nu}
    + \rk_{\mu} \rk_{\nu} \right\}}
        {\bigl[ \rk^2- (y\rqq+ x\rp_2)^2 -
          (x\tilde{m}^2 -y\rp_1^2 -(x-y)\rp_2^2) \bigr]^3} \ ,
\eeq
where $\rqq = \rp_1-\rp_2$, $\tilde{m}^2 = m^2-i0$, and the identity (\ref{si3a}) has
been taken into account. Shifting the integration variable $y \to xy$ yields
\beq                                            \label{ap10}
\left\{ I, I_{\mu}, I_{\mu \nu} \right\} =
       \frac{32\pi^2}{i} \int_0^1 dy \int_0^1 dx x \int \frac{d^D\rk}{(2\pi)^D}
\frac{\left\{1;\,  x(y\rqq+\rp_2)_{\mu};\,  x^2(y\rqq+\rp_2)_{\mu} (y\rqq+\rp_2)_{\nu}
    + \rk_{\mu} \rk_{\nu} \right\}}
        {\bigl[ \rk^2- x^2(y\rqq+ \rp_2)^2 -
          x(\tilde{m}^2 -y\rp_1^2 -(1-y)\rp_2^2) \bigr]^3} \ .
\eeq
Now we separate the integral $I_{\mu\nu}$ into two parts, $I_{\mu\nu} = I^a_{\mu\nu}+
I^b_{\mu\nu}$. Here, $I^b_{\mu\nu}$ corresponds to the part of $I_{\mu\nu}$ with
$\rk_{\mu}\rk_{\nu}$ in the numerator, and $I^a_{\mu\nu}$ is the remainder.
Integration over $\rk$ yields ($D = 4-\epsilon$),
\beqn                                            \label{ap11}
\left\{ I, I_{\mu}, I^a_{\mu \nu} \right\} &=&
       -\Gamma(1+\epsilon/2) (4\pi)^{\epsilon/2}
       \int_0^1 dy \int_0^1 dx
\frac{
    \left\{1;\,  x(y\rqq+\rp_2)_{\mu};\,  x^2(y\rqq+\rp_2)_{\mu} (y\rqq+\rp_2)_{\nu} \right\}}
        {x^{\epsilon/2} N^{1+\epsilon/2}}
        \ ,
\eeqn
\beq                                            \label{ap12}
I^b_{\mu \nu} =
       \frac2{\epsilon} \Gamma(1+\epsilon/2) (4\pi)^{\epsilon/2} \frac{g_{\mu \nu}}{2}
       \int_0^1 dy \int_0^1 dx
\frac{ x^{1-\epsilon/2}}
        {N^{\epsilon/2}} \ ,
\eeq
where $ N = x(y\rp_1+ (1-y)\rp_2)^2  + \tilde{m}^2 -y\rp_1^2 -(1-y)\rp_2^2$. In the
limit $\epsilon \to 0$, we have
\beqn                                            \label{ap13}
\left\{ I, I_{\mu}, I^a_{\mu \nu} \right\} &=&
       - \int_0^1 dy \int_0^1 dx
\frac{
    \left\{1;\,  x(y\rqq+\rp_2)_{\mu};\,  x^2(y\rqq+\rp_2)_{\mu} (y\rqq+\rp_2)_{\nu} \right\}}
        {N} + O(\epsilon)
        \ ,
\eeqn
\beq                                            \label{ap14}
I^b_{\mu \nu} =
    \frac{g_{\mu \nu}}{4} \Delta_{\epsilon}  - \frac{g_{\mu \nu}}{2}
       \int_0^1 dy \int_0^1 dx\, x\, \ln N  + O(\epsilon)
        \ ,
\eeq
where $\Delta_{\epsilon} = 2/\epsilon -\gamma_E+ \ln 4\pi$.

Obviously, the denominator $N$ is a quadratic polynomial with respect to $\omega$. Let
us find its roots. In kinematics ``a'', we have:
\beq                                            \label{ap16}
N = -(1-x)(\vare_a-\omega)^2  + \tilde{m}^2 + B^2 \ ,
\eeq
where $B^2 = xy(1-y)\bfq^2+ (1-x)y\bfp_1^2 + (1-x)(1-y)\bfp_2^2$, $B^2 \ge 0$. We
write $N$ as
\beq                                            \label{ap17}
N = (1-x)(\omega_{+}-\omega-i0)(\omega_{-}+\omega-i0) \ ,
\eeq
where
\beq                                            \label{ap18}
\omega_{\pm} = \sqrt{\displaystyle
        \frac{m^2+ B^2}{1-x}} \pm \vare_a \ .
\eeq
As can easily be seen, $\omega_{\pm} \ge \omega_{\pm}^0 = m\pm \vare_a$.

In kinematics ``b'', we have analogously,
\beq                                            \label{ap19}
N = y(1-xy)(\omega_{+}-\omega-i0)(\omega_{-}+\omega-i0) \ ,
\eeq
where
\beq                                            \label{ap20}
\omega_{\pm} = \frac{1-x}{1-xy} \left\{\sqrt{\vare_a^2+
        \frac{1-xy}{y(1-x)^2} \bigl( m^2-(1-x)\vare_a^2+ B^2 \bigr)}
                \pm \vare_a \right\} \ .
\eeq
Again, one can show that $\omega_{\pm} \ge \omega_{\pm}^0 = m\pm \vare_a$.

So, we find that $\Gamma^{\sigma}(\vare_a-\omega,\vare_a-\omega)$ and
$\Gamma^{\sigma}(\vare_a-\omega,\vare_a)$ are analytic functions of $\omega$ in the
complex plane with the branch cuts $[m+\vare_a-i0, \infty-i0)$ and $[-m+\vare_a+i0,
-\infty+i0)$.

%

\newpage
\begin{table}
 \caption{The individual contributions to $\Delta E^{N1}_P$, in a.u.
\label{Table_dEN1}}
\vspace*{0.5cm}
\begin{tabular}{cr@{.}lr@{.}lr@{.}lr@{.}l}
$Z$ & \multicolumn{2}{c}{$\Delta E^{N1}_{P,\rm infr}$} &
                \multicolumn{2}{c}{$\Delta E^{N1}_{P,\rm pole}$} &
                        \multicolumn{2}{c}{$\Delta E^{N1}_{P,\rm Im}$} &
                            \multicolumn{2}{c}{ Total }\\
        \hline
83 &  $-$0&03419  &  0&00480  & $-$0&03951  & $-$0&06890(20) \\
90 &  $-$0&03484  &  0&00430  & $-$0&03780  & $-$0&06834(20) \\
92 &  $-$0&03489  &  0&00438  & $-$0&03737  & $-$0&06788(20)
\end{tabular}
\end{table}

\begin{table}
 \caption{The individual contributions to $\Delta E^{N2}_P$, in a.u.
\label{Table_dEN2}}
\vspace*{0.5cm}
\begin{tabular}{cr@{.}lr@{.}lr@{.}lr@{.}l}
$Z$ & \multicolumn{2}{c}{$\Delta E^{N2}_{P,\rm infr}$} &
                \multicolumn{2}{c}{$\Delta E^{N2}_{P,\rm pole}$} &
                        \multicolumn{2}{c}{$\Delta E^{N2}_{P,\rm Im}$} &
                                          \multicolumn{2}{c}{Total} \\
        \hline
83 &  0&06563  &  $-$0&02127  &  0&02178  &  0&06614(20) \\
90 &  0&07200  &  $-$0&02675  &  0&02460  &  0&06985(20) \\
92 &  0&07403  &  $-$0&02881  &  0&02570  &  0&07092(20)
\end{tabular}
\end{table}

\begin{table}
 \caption{The individual contributions to $\Delta E^{O1}_P$, in a.u.
\label{Table_dEO1}}
\vspace*{0.5cm}
\begin{tabular}{cr@{.}lr@{.}lr@{.}l}
$Z$ & \multicolumn{2}{c}{$\Delta E^{O1}_{P,\rm pole}$} &
                        \multicolumn{2}{c}{$\Delta E^{O1}_{P,\rm Im}$} &
                             \multicolumn{2}{c}{Total} \\
        \hline
83 &  $-$0&03602  &  0&02496  &  $-$0&01106(15) \\
90 &  $-$0&03599  &  0&02417  &  $-$0&01182(15) \\
92 &  $-$0&03605  &  0&02392  &  $-$0&01213(15)
\end{tabular}
\end{table}

\begin{table}
 \caption{Finite parts of the individual contributions to $\Delta E$, in a.u.
\label{Table_dE}}
\vspace*{0.5cm}
\begin{tabular}{cr@{.}lr@{.}lr@{.}lr@{.}l}
$Z$ & \multicolumn{2}{c}{$\Delta E^{N1}$} &
                  \multicolumn{2}{c}{$\Delta E^{N2}$} &
                        \multicolumn{2}{c}{2$\Delta E^{O1}$} &
                                \multicolumn{2}{c}{Total} \\
        \hline
83 & $-$0&06890 &   0&06614 &   $-$0&02212  &  $-$0&02488(40) \\
90 & $-$0&06834 &   0&06985 &   $-$0&02364  &  $-$0&02213(40) \\
92 & $-$0&06788 &   0&07092 &   $-$0&02426  &  $-$0&02122(40)
\end{tabular}
\end{table}

%
\newpage
\begin{table}
\caption{The ground-state Lamb shift in ${}^{238}{\rm U}^{91+}$, in eV.
\label{uranium}}
\begin{tabular}{llr@{.}lc}
Finite nuclear size &                     &  198 & 81(38)  &  \\
First-order &  self-energy                &  355 & 05      & \cite{MohrSoff93} \\
            & vacuum polarization         &$-$88 & 60      & \cite{SoffMohr88}  \\
Second-order & SESE  (a, irred.)          & $-$0 & 97      & \cite{Mitrushenkov95}\\
             & SESE  (a, red.) (b,c)      & $-$0 & 90(1)   &This work + \cite{Mallampalli98} \\
             & VPVP  (d)                  & $-$0 & 22      & \cite{Persson96a,Beier97}\\
             & VPVP  (e,f)                & $-$0 & 75(30)  & \cite{Beier88,Schneider93,Plunien98}   \\
             & SEVP  (g-i)                &    1 & 12      & \cite{Persson96a}\\
             & S(VP)E (k)                 &    0 & 13(6)   & \cite{Persson96a,Mallampalli96} \\
             & Total  (a-k)               & $-$1 & 60(31)  &  \\
Nuclear recoil &                          &    0 & 46      & \cite{Shabaev98}\\
Nuclear polarization &                    & $-$0 & 20(10)  & \cite{Plunien95,Nefiodov96,Yamanaka01}\\
\hline
Lamb shift (theory)  &                    &   463 & 93(50)&   \\
Lamb shift (experiment) &                 &   468 &    $\pm$ 13. & \cite{Stoehlker00}
\end{tabular}
\end{table}

\begin{table}
\caption{The $2p_{1/2}$-$2s$ transition energy in Li-like ${}^{238}$U, in eV. The
first quoted error in the total theoretical prediction arises from the uncertainty due
to the finite nuclear size effect and due to higher-order electron correlations. The
second quoted error corresponds to the uncertainty of the second-order one-electron
QED effects. \label{Liuranium} }
\begin{tabular}{llr@{.}lc}
\multicolumn{2}{l}{Transition energy without second-order one-electron QED effects}
                                          & 280 & 48(11) & \cite{Yerokhin01} \\
One-electron second-order & SESE (a-c)    &   0 & 23(20) & This work\\
             & VPVP  (d)                  &   0 & 04     & \cite{Persson93,Persson96a,Beier97}\\
             & VPVP  (e,f)                &   0 & 10(5)  & \cite{Beier88,Schneider93,Plunien98} \\
             & SEVP  (g-i)                &$-$0 & 19     &  \cite{Lindgren93,Persson96a}\\
             & S(VP)E (k)                 &$-$0 & 02(1)  &  \cite{Persson96a}\\
             & Total  (a-k)               &   0 & 15(21) & \\
\hline
Transition energy (theory)  &             & 280 & 64(11)(21) &   \\
Transition energy (experiment) &          & 280 & 59(10)    & \cite{Schweppe91}
\end{tabular}
\end{table}

\newpage


\end{document}